\documentclass{aastex}
\usepackage{emulateapj5}
\usepackage{apjfonts}
\usepackage{epsf}

\slugcomment{Received 2005 May 5; accepted 2005 July 28}
\journalinfo{The Astrophysical Journal Letters in Press}
\shorttitle{Feedback from Captures of Red Giant Stars to Hot Gas}
\shortauthors{WANG \& HU}

\newcommand{\ce}{\ifmmode {\cal E} \else ${\cal E}$\ \fi}

\def\sunm{M_{\odot}}

\newcommand{\kms}{\ifmmode {\rm km\ s}^{-1} \else km s$^{-1}$\ \fi}
\newcommand{\ergs}{\ifmmode {\rm erg\ s}^{-1} \else erg s$^{-1}$\ \fi}
\newcommand{\tes}{\ifmmode \tau_{\rm es} \else $\tau_{\rm es}$\ \fi}
\newcommand{\tk}{\ifmmode \tau_{\rm K} \else $\tau_{\rm K}$\ \fi}
\newcommand{\vfwhm}{\ifmmode V_{\mbox{\tiny FWHM}} \else
            $V_{\mbox{\tiny FWHM}}$\fi}
\newcommand{\msun}{\ifmmode M_{\odot} \else $M_{\odot}$\ \fi}
\newcommand{\afe}{\ifmmode {\mathcal A_{\rm Fe}} \else${\mathcal
    A_{\rm Fe}}$\ \fi}

\newcommand{\ledd}{\ifmmode L_{\rm Edd} \else $L_{\rm Edd}$\ \fi}
\newcommand{\lx}{\ifmmode L_{\rm 2-10keV} \else  $L_{\rm
    2-10keV}$\ \fi}
\newcommand{\hb}{\ifmmode H\beta \else H$\beta$\ \fi}
\newcommand{\mbh}{\ifmmode M_{\rm BH}  \else $M_{\rm BH}$\ \fi}

\received{????}
\begin{document}

\title{Captures of Red Giant Stars by Black Holes in Elliptical
  Galaxies: \\
Feedback to the Hot Gas}

\author{Jian-Min Wang\altaffilmark{1} and  
        Chen Hu\altaffilmark{2}}

\altaffiltext{1}{Key Laboratory for Particle Astrophysics, Institute
  of High Energy Physics,
Chinese Academy of Sciences, Beijing 100049, China;
wangjm@mail.ihep.ac.cn} 

\altaffiltext{2}{National Astronomical Observatories, Chinese Academy
  of Sciences, Beijing 100012, China}

\begin{abstract}
The highly disturbed hot gas in elliptical galaxies, 
as revealed in many 
{\em Chandra} X-ray images, implies a source of energy in the 
galactic nucleus.
In some elliptical galaxies 
faint X-ray ``ghost'' cavities appear without corresponding radio lobes.
It has been suggested that ghost cavities are caused by 
short-lived activity with a timescale of $\sim 10^3-10^4$ years,
but this is  
difficult to understand within the popular paradigm of 
active galactic nuclei.
We suggest an episode model for ghost cavities, invoking
captures of red giant stars by the black hole 
located at the center of the elliptical galaxies at a typical rate
of $10^{-5}$yr$^{-1}$ per galaxy. 
The accretion of tidally disrupted red giant stars onto the black
hole powers activity in a timescale of 
a few years. The total energy channeled into the jet/outflow during
the cooling time of the hot gas is 
$\sim 10^{56}$ erg, which is the typical work required to form the
observed cavities. 
In this scenario, the faint cavities are
produced by the feedback following accretion of the debris of the captured
red giant stars onto the black holes. 
We apply the present model to several elliptical galaxies and  
find that it can explain the formation of the ghost
cavities. 
This model can be tested in the future by 
comparisons between radio and X-ray observations. 

\end{abstract}

\section{INTRODUCTION}
Black holes located in the centers of galaxies inevitably capture stars 
(Hills 1975; Rees 1988) 
if the tidal energy dissipated in the stellar envelope exceeds
its orbital energy with respect
to the black hole. For 
a main sequence star of mass $M_*$ and radius $R_*$, the captures 
occur at a tidal 
radius, $R_{\rm T}\approx R_*\left(M_{\rm BH}/M_*\right)^{1/3}$, which
is beyond the last stable 
orbit of a black hole with mass $M_{\rm BH}\le 2\times 10^8\sunm$. 
Such capture events occur 
every $\sim 10^4$ years (Rees 1988). 
A fraction of the gas released by the disruption
of the stars 
remains bound to the black hole and eventually 
will be accreted.  This process would create bright flares
in  optical, UV and soft X-ray %as it accrets onto the black hole 
with a timescale 
of a few years (Rees 1988; Cannizzo et al. 1990; Loeb \& Ulmer
1997).
{\em Chandra} observations with high spatial resolution 
of the  nonactive galaxy
RX J1242.6-1119A
show striking evidence for the capture process 
(Komossa et al. 2004) and a growing body of evidence provides further 
support for such captures (Halpern et al. 2004).

Larger black holes, $M_{\rm BH}\ge 2\times 10^8\sunm$, can 
entirely swallow a captured main sequence star 
with very little radiative losses. 
However, giant stars can be tidally
disrupted by the black hole 
beyond the last stable orbit since their densities are much lower than 
main sequence stars (Syer \& 
Ulmer 1999). 
Therefore the debris of captured red giants will significantly
radiate its gravitational binding 
energy as it accrets onto the black hole. Although the energy released
by each red giant capture 
is small, the total energy released during the lifetime of the
galaxy is quite large, extending to 
$> 10^{56}$ erg if some of the released energy can be piled up 
before the gas cools. 
An interesting question arises: Can the capture of red
giant stars significantly influence the hot gas in these galaxies? 

It is well-known that elliptical galaxies contain significant
amounts of hot gas originally found 
by {\em Einstein} (Forman et al. 1985; see a review of Mathews \&
Brighenti 2003). It has been suggested
that the hot gas can be disturbed in two ways: (1) 
by internal processes 
such as nuclear outbursts; and (2) externally by 
galaxy-galaxy and galaxy-cluster interactions.
In this Letter we are only concerned with the first case. 
When powerful radio jets interact strongly with the hot gas
very clear features of cavities, hot spots etc. should appear, as in 
MS 0735.6+7421 (McNamara et al. 2005), where the radio lobes fill the
X-ray cavities. However, faint ``ghost'' X-ray cavities are also observed 
in the absence of powerful radio sources. 
For example, NGC 4636, a well-known non-active galaxy without radio lobes,
has an extensive, highly disturbed X-ray halo 
containing ghost cavities (Jones et al. 2002; Ohto et
al. 2003). This strongly implies a very short timescale for the 
activity of the black hole in this galaxy, 
$10^3 - 10^4$ years. Observations typically constrain the lifetime
of active galactic nuclei to 
$10^6 - 10^8$ years (Martini 2004), but such long-term activity cannot 
apply to NGC 4636. 
What powers such short-lived activity on galactic scales? 

In this paper we focus on the capture of red giant stars by black
holes of 
mass $> 2\times 10^8M_{\odot}$. We show that the accretion of 
stellar debris onto 
the black holes produces an energy feedback 
into the galactic hot gas that can explain 
the formation of X-ray ghost cavities.

\section{Red Giant Star Capture Process and Feedback}

\subsection{Star-capture rates of the black holes}
In their study of the capture rate of red giant stars by 
massive galactic black holes, 
Syer \& Ulmer (1999) made the following assumptions:
1) a virialized star cluster;
2) a simple model of stellar evolution;
3) a Salpeter mass function for the stars; and 
4) the maximum radius of red giant stars is 
determined by collisions with main sequence stars. 
For a Salpeter IMF a fraction 0.05 of stars are red giants. 
The maximum radius of the red giant stars, 
$200R_{\odot}$, is attained only during a 
short phase ($10\%$) of their evolutionary lifetime. 
The mean radius
of red giant stars is ${\bar R}_*=12R_{\odot}$.
The average tidal radius under a Newtonian potential is given by
\begin{equation}
\frac{\bar{R}_{\rm
    T}}{R_S}=\frac{\bar{R}_*}{R_S}\left(\frac{\mbh}{M_*}\right)^{1/3}
         =26.3~m_*^{1/3}M_8^{-2/3}r_*,
\end{equation}
where $R_S=G\mbh/c^2$ is the Schwarzschild
radius of the black hole, 
$r_*={\bar
  R}_*/12R_{\odot}$ and $m_*=M_*/\sunm$ are 
respectively the radius and mass of the captured star, 
and $M_8=\mbh/10^8\sunm$ is the mass of the black hole.

Since the radius of the red giant star is much larger than that of main
sequence stars, the loss 
cone will be enlarged by a factor of $R_{\rm max}/R_{\odot}$. A
detailed calculation of the red giant capture 
rate by Syer \& Ulmer (1999) depends on a large
number of parameters: star density 
profile, total number of stars, Salpeter mass function, the maximum
radius of the red giant stars,
the black hole mass and the time-dependent radius of the red giant
stars. However, a simplified version of
the capture rate can be obtained by fitting the data in Syer \& Ulmer
(1999),
\begin{equation}
\dot{N}_{\rm RG}\approx
10^{-5}\left(\frac{\mbh}{10^8\sunm}\right)^{0.35}~{\rm yr^{-1}},
\end{equation}
where other parameters are fixed.
This approximation is reasonably good for more massive galaxies
where $M_{\rm BH}\ge 10^8M_{\odot}$.

\subsection{Impulsive Feedback Heating}
The detailed capture process and relevant radiation from the bound gas
have been studied by several 
authors (e.g. Rees 1988; Cannizzo et al. 1990; Loeb \& Ulmer 1997;
Ulmer 1999). We follow treatments
in Loeb \& Ulmer (1997) and Ulmer (1999).
Once a red giant star is tidally disrupted, a fraction $\xi$ of its 
gas remains bound to the 
BH. This gas will return to the pericenter after a time $t_{\rm min}$
(Ulmer 1999)
\begin{equation}
t_{\rm min}\approx 45.7~ r_*^{3/2}m_*^{-1}M_8^{1/2}
                       \left(\frac{R_p}{{\bar{R}_{\rm
                             T}}}\right)^3~{\rm yr},
\end{equation}
where $R_p$ is the pericenter radius of the captured star ($R_p\sim
{\bar R}_{\rm T}$). 
The gas enters a circular orbit around the 
BH and forms an accretion disk within a timescale  
\begin{equation}
t_{\rm cir}=n_{\rm orb}t_{\rm min}\approx 91.5\left(\frac{n_{\rm
    orb}}{2}\right)                       
                                             \left(\frac{t_{\rm
                                                 min}}{45~{\rm
                                                 yr}}\right)~{\rm yr},
\end{equation}
where $n_{\rm orb}$ is a small number of orbits necessary for circularization
(Ulmer 1999). Subsequently, a thick
torus or thin disk will form, depending 
on accretion timescale $t_{\rm acc}$ and radiation timescale $t_{\rm
  rad}$
(the time to radiate all of the energy released by the debris at the 
Eddington luminosity; Loeb \& Ulmer 
1997). If $t_{\rm acc}\gtrsim t_{\rm rad}$, a thin disk will form,
otherwise, a torus.  
Using the parameterized viscosity through the relation
$\eta=\alpha P_{\rm rad}\Omega_{\rm K}^{-1}$, where $\alpha$ is the
viscous constant and the Keplerian 
velocity $\Omega_{\rm K}=(G\mbh/R^3)^{1/2}$(Shakura \& Sunyaev 1973),
we find 
the debris of the captured star will be swallowed by the black hole
within a time  
\begin{equation}
t_{\rm acc}\approx\frac{\rho_{\rm gas}{\bar{R}}_{\rm T}^2}{\eta}
           \approx 0.8~M_8\alpha_{-2}^{-1}\left(\frac{r_{\rm
               T}}{25}\right)^{3/2}~{\rm yr},
\end{equation}
where 
$\alpha_{-2}=\alpha/10^{-2}$ and $r_{\rm T}=\bar{R}_{\rm T}/R_S$
(Eq. 20 in Loeb \& Ulmer 1997). The radiation 
time is
\begin{equation}
t_{\rm rad}=\frac{\xi\epsilon m_*\sunm c^2}{L_{\rm Edd}}
            \approx 0.21~\xi_{0.5}m_*M_8^{-1}\epsilon_{0.1}~{\rm yr},
\end{equation}
where $\epsilon_{0.1}=\epsilon/0.1$ is the the accretion efficiency
and $\xi_{0.5}=\xi/0.5$. We find $t_{\rm acc}>t_{\rm rad}$, 
therefore a
thin disk forms.
The structure of the disk is determined by the Eddington ratio 
defined by $\dot{m}=\dot{M}/\dot{M}_{\rm Edd}$ where $\dot{M}_{\rm
  Edd}=2.2\epsilon_{0.1}^{-1}M_8~\sunm$/yr, therefore 
\begin{equation}
\dot{m}\approx \frac{\xi M_*}{t_{\rm acc}\dot{M}_{\rm Edd}}
       \approx 0.3~ \xi_{0.5}m_*M_8^{-1}.
\end{equation}
The disk formed by the debris corresponds to the inner region of the
standard disk unless the accretion 
timescale is much shorter than $t_{\rm rad} \sim 0.21$yr. From the
standard accretion disk model (Shakura 
\& Sunyaev 1973), we find $P_{\rm rad}/P_{\rm gas}=
3.9\times 10^2\left(\alpha_{-2}M_8\right)^{1/4}\left(r_{\rm
  T}/25\right)^{-21/8}\dot{m}_{-1}^2\gg 1$, 
where $\dot{m}_{-1}=\dot{m}/0.1$, so the radiation pressure
dominates the gas in the debris disk.

It has been suggested that most of the black holes in the universe are
rapidly spinning (Elvis et al. 
2002; Volonteri et al. 2005). 
The energy channeled into an outflow by the Blandford \& Znajek (1977)
mechanism from a radiation-pressure-dominated 
disk has been studied by Ghosh \& Abramowicz (1997),  who show the
ratio of the jet power to the accretion
luminosity is $\epsilon_{\rm j}=3.2\times 10^{-2}\dot{m}_{-1}^{-1}$
(equation 15 in Ghosh \& Abramowicz 1997).
This is consistent with $\epsilon_{\rm j}\propto \dot{m}^{-0.65}$ from
blazars statistics (Wang et 
al. 2004).  The kinetic energy of the jet during the entire accretion
is given by 
%
%\begin{eqnarray}
\begin{equation}
E_{\rm K}=\epsilon_{\rm j}E_{\rm acc}=2.4\times
10^{52}~\epsilon_{0.42}M_8~{\rm erg},
%\end{eqnarray}
\end{equation}
where $\epsilon=0.42\epsilon_{0.42}$ is the accretion efficiency for a 
black hole with spin $a=1$. We 
note that the dependence of the kinetic energy on the 
black hole mass is determined by $\epsilon_{\rm j}$.
 
The interaction between the jet/outflow and the hot gas involves 
many micro-physical processes. 
Detailed calculations of the heating processes due to these 
interactions are
beyond the scope of the present paper. However, we can easily estimate
the location of the 
cavities from the jet/outflow energy. 
The outflow is sharply decelerated when the thermal energy of swept up and
shocked ambient gas equals the initial kinetic energy of the outflow.
Therefore $M_{\rm j}v^2\approx \pi \Theta^2R^3n_em_pc_{\rm s}^2/3$, 
where $\Theta$ is the opening angle of the outflow, $c_s$ is the sound
speed, $v$ is the velocity of the jet/outflow, $m_p$ is the proton mass
and $M_{\rm j}$ is the mass of the outflow. The location of the
cavities is then 
$R\approx 4.2
(\Theta/0.05)^{2/3}n_{-2}^{-1/3}(v_{0.1}/c_{s,2})^{2/3}\left(\xi_{0.5}m_*\right)^{1/3}$~
kpc, 
where $c_{s,2}=c_s/100~{\rm km~s^{-1}}$, $v_{0.1}=v/0.1c$ and $c$ is
the light speed. 
This radius is consistent with {\em Chandra} observations.
Without specifying mechanisms, the time for the outflow from 
the black hole to dissipate its kinetic energy is 
\begin{equation}
t_{\rm heating}=1.7\times 10^5 R_{\rm 5kpc}v_{0.1}^{-1}~{\rm yr},
\end{equation}
where $R_{\rm 5kpc}=R/5{\rm kpc}$.
Cavities of heated gas expand until pressure equilibrium is reached with
their surroundings, $n_cT_c=n_hT_h$. From the brightness contrast of observed
X-ray images the cavities are characterized by $fn_c^2 T_c^{1/2}=n_h^2T_h^{1/2}$ where $f>1$, 
and the subscripts of $c$ and $h$ refer to cavity and hot gas, respectively. 
At this time, $T_c>T_h$ and $n_c<n_h$ hold. The equilibrium continues until the cavities cool '
by bremsstrahlung. The lifetime of X-ray cavities is thus determined by the cooling time, 
\begin{equation}
t_{\rm ff}=1.8\times 10^9T_7^{1/2}n_{-2}^{-1}~{\rm yr},
\end{equation}
where $T_7=T/10^7{\rm K}$. Since the cooling timescale is much longer
than the heating, cavities
formed by successive stellar captures will merge, i.e., the
energy injected in the cavities 
will pile up. 
The total kinetic energy released by 
the tidal disruption of a series of stellar captures by the 
black hole is given by
\begin{equation}
E_{\rm K}^{\rm tot}=\sum_{i=1}^{n}E_{\rm K}^i;~~~{\rm and}~~~
n=\dot{N}\min(t_{\rm G},t_{\rm ff}),
\end{equation}
where $n$ is the total number of star-capture events during time
$\min(t_{\rm ff},t_{\rm G}$), 
$t_{\rm G}$ is the lifetime of galaxies, and $E_{\rm K}^i$ is 
the kinetic energy released in event $i$. When $t_{\rm ff}\le t_{\rm
  G}$, some of the cavities
produced by the outflow disappear. The radius of the cavity can be
simply estimated by $P\Delta V=E_{\rm K}$,
where $P$ is the pressure of the hot gas and $\Delta V$ is the volume
of the cavities. We have 
\begin{equation}
\Delta R\approx \left(\frac{3E_{\rm K}^{\rm tot}}{4\pi P}\right)^{1/3}
        \approx 3.87~E_{56}^{1/3}n_{-2}^{-1/3}T_7^{-1/3}~{\rm kpc},
\end{equation}
where $E_{56}=E_{\rm K}^{\rm tot}/10^{56}~{\rm erg}$.

\begin{table*}[t]
\begin{center}
\centerline{\sc Table 1.}
\footnotesize
\centerline{\sc Star Captures and Cavities in Elliptical Galaxies}
\vglue 0.1cm
\begin{tabular}{lcccccccccc}\hline\hline
Name & $\sigma$& Ref. & Age & Ref. & $\log\mbh$ & $\dot{N}(\times
10^{-5})$ & 
$E_{\rm K}^{\rm tot}(\times 10^{56})$ & $P\Delta V(\times 10^{56})$ &
$R$ & Ref. \\
         & (km~{$\rm s^{-1}$})  & & (Gyr) & & ($M_{\odot}$) &
(yr$^{-1}$) & (erg)& (erg)& (kpc) &   \\     
(1)      &  (2)    &(3)& (4)& (5) & (6) &(7) & (8)& (9) & (10) & (11)
\\ \hline
NGC 708 (A 262)    & 241 & 1 &    &   & 8.46 & 1.44  &   17.8 & 4.25
& 6.5 & 3,5 \\
NGC 4472    & 273 & 2 & 8.5& 7 & 8.67 & 6.31  &    128 & 0.013 & 3.6 &
4 \\ 
NGC 4636    & 180 & 2 &    &   & 7.95 & 0.447 &   1.71 & 0.24  & 5.1 &
6 \\ \hline

\end{tabular}
\parbox{6.0in}
{\baselineskip 9pt
\noindent
{\sc References:} 
1. McElroy (1995);
2. Merritt \& Ferrarese (2001);
3. Blanton, et al. (2003);
4. Biller, et al. (2004);
5. Birzan, et al. (2004);
6. Ohto, et al. (2003);
7. Terlevich \& Forbes (2002)
}
\end{center}
\vglue -1.0cm
\end{table*}
\normalsize

We note that $\max(t_{\rm rad}, t_{\rm cir}, t_{\rm acc})\ll
1/\dot{N}_{\rm RG}\sim 10^5~{\rm yr}$.
This means that the feedback to the hot gas is impulsively episodic, not
continuous. 
In summary this model involves two initial steps: 1)
a red giant star is captured and 2)
formation of an accretion disk from its debris. 
The first step depends on stellar
evolution in 
the galaxy and the second involves hydrodynamical processes 
that eventually produce 
jets/outflows. We stress here that these assumptions may be
improved and refined in the future by more detailed 
comparisons with observations, 
but currently available data is sufficient to 
support the viability of this model for cavity 
production.

\section{Applications and Discussions}
Table 1 lists properties of hot gas cavities in three
elliptical galaxies observed with {\em Chandra}.
These ``ghost'' X-ray cavities are in relatively radio-quiet galaxies
and are thus not caused by powerful jets as observed in 
radio-loud galaxies and quasars. %In addition, 
%the required timescale of nuclear activity for the cavities can be
%simply estimated by
%$t\approx P\Delta V/L_{\rm BZ}\approx 1.6\times
%10^4\dot{m}_{-1}^{-1}M_8^{-1}$ years for 
%$P\Delta V=10^{56}$ erg and Kerr black holes. This timescale is much
%shorter than lifetimes of any 
%known kind of active galactic nuclei (Martini 2004). 
Col (1) gives the name of the elliptical; Col (2) lists the 
stellar velocity dispersion $\sigma$ in the galaxies; 
Col (3) lists the references for $\sigma$; Col (4) provides the lifetime
of the galaxy if it has been estimated in the published references
given in Col (5); Col (6) provides the black hole mass estimated from $\log
\mbh/\sunm=8.13+4.02\log\left(\sigma/200\kms\right)$
(Tremaine et al. 2002); Col (7) provides the capture rates taken from  
Syer \& Ulmer (1999) or estimated from equation (2); Col (8) gives the
total kinetic
energy released by red giant captures 
during the galactic lifetime $t_{\rm G}$ or cooling timescale based on
equation (11);  
Cols (9) and (10) show respectively the work done to 
created the cavities and the distance of the cavities from the
center
estimated from {\em Chandra} images; Col (11) gives the
relevant references.

{\em NGC 708} is a cD galaxy at the center of cluster A 262. It has a 
fairly faint double-lobed FR I
radio morphology with a radio power $P_{1.4}=4.7\times
10^{22}~{\rm W~ Hz^{-1}}$ at 1.4 GHz 
(Blanton et al. 2004), indicating that this galaxy has not undergone a
long-term radio-loud phase. 
Its VLA image is coincident with the X-ray cavities (Blanton et
al. 2004). Table 1 shows that the 
feedback energy released by accretion onto the black hole of
$10^{8.46}\sunm$ is sufficient
to power the formation of the ``ghost'' cavities. {\em NGC 4472} is a
giant elliptical galaxy in the 
Virgo cluster with a faint radio jet/lobe similar to NGC 708. 
The red giant capture rate estimated by 
Syer \& Ulmer (1999) is $10^{-4.2}$ ${\rm yr ^{-1}}$, which is 
relatively high. The 
feedback energy given in Table 1 is much larger than the work 
required to form the 
X-ray cavities. The giant 
elliptical {\em NGC 4636}, which is on the outskirts of the Virgo cluster,
shows no evidence of radio jet/lobes. 
Ohto et al. (2003; see also O'Sullivan et al. 2005 for more details)
revisited the {\em Chandra} ACIS data and showed that there are
symmetric X-ray cavities along the
south-west to north-east direction. They argue that the
cavities are caused by the nuclear 
activities with a timescale of $\sim 10^3$ yr. 
Ongoing captures of red giant stars by the central black hole in 
NGC 4636 can explain the X-ray cavities in the surrounding gas. 

%\section{Discussions}
%If the black holes are only slowly spinning, the Blandford-Payne wind
%developed from a magnetized 
%accretion disk would be expected to have feedback to the hot gas. 
%According to Meier (2001), the kinetic power of the Blandford-Payne
%wind
%is $L_{\rm
%jet}=10^{40.8}\alpha_{-2}^{-1/10}M_8^{9/10}\dot{m}_{-1}^{6/5}$\ergs
%for a Schwarzschild black
%hole Shakura-Sunyaev disk. The timescale for the typical cavities is
%then 
%$t\approx P\Delta V/L_{\rm jet}=3.0\times
%10^7\alpha_{-2}^{1/10}M_8^{-8/9}\dot{m}_{-1}^{-6/5}$yr.
%Such a long activity in elliptical galaxies with ghost X-ray cavities
%has not been found. The Blandford-Payne
%wind is excluded as energy source powering X-ray cavities.

The detailed hot gas morphologies indicate that the cavity formation 
process is more complicated than 
the model presented here.   Our red giant capture rates are 
calculated by simplifying several
important processes in Syer \& Ulmer (1999) such as for example 
our assumption of a simple stellar 
evolution model with a constant lifetime 
for red giant stars ($t_{\rm RG}\sim 7\times 10^8$ yr). We take the
average radius of the red giant 
during its lifetime, 
while in reality the capture process will depend in detail 
on the evolutionary phase of the giant star. 
These simplifying assumptions
lead to uncertainties when 
applied to individual galaxies. 
A key test to support or verify the capture model would be 
observations showing radio jets that appear 
simultaneous with X-ray flares in galaxies. 

\section{Conclusions}
The highly disturbed hot gas in elliptical galaxies implies a short
active
history of the black holes. 
Captures of red giant
stars and the feedback from their debris can explain the short 
activity timescales that create cavities
in the hot gas and we show quantitatively that such a process may work in
elliptical galaxies.
An advantage of the present model is that red giant captures are 
a natural process 
that can trigger short-lived activity near the black hole. 
In particular, 
the accumulation of feedback energy resulting from successive red giant 
captures can explain the cavities observed in elliptical galaxies. 
This can be done in the absence of powerful, long-lived ($\sim 10^8$ year) 
radio emission. Further, more detailed studies of the red giant 
capture process are likely to suggest additional 
observational consequences. 

\acknowledgements{The authors express their thanks to an anonymous
referee for detailed reports improving the paper. J.M.W\@. is grateful to W. G. 
Mathews for a careful reading of the manuscript, a number of comments and 
encouragements. S. Komossa is acknowledged for useful comments on star captures.
He thanks the supports from a Grant for Distinguished Young Scientist
from NSFC, NSFC-10233030, the
Hundred Talent Program of CAS and the 973 project.}

\end{document}